\documentclass[prl,aps,showpacs,twocolumn]{revtex4}
\usepackage{graphicx}
\usepackage{amsmath}
\usepackage{amsfonts}
\usepackage{amssymb}

\begin{document}

\title{Polymorphic Dynamics of Microtubules}
\author{Herv\'{e} Mohrbach$^{1}$, Albert Johner$^{2}$ and Igor M. Kuli\'{c}$^{2}$}
\email[Email: ]{kulic@unistra.fr}
\date{\today}
\begin{abstract}
Starting from the hypothesis that the tubulin dimer is a
conformationally bistable molecule - fluctuating between a curved
and a straight configuration at room temperature - we develop a
model for polymorphic dynamics of the microtubule lattice. We show
that tubulin bistability consistently explains unusual dynamic
fluctuations, the apparent length-stiffness relation of grafted
microtubules and the curved-helical appearance of microtubules in
general. Analyzing experimental data we conclude that taxol
stabilized microtubules exist in highly cooperative yet strongly
fluctuating helical states. When clamped by the end the
microtubule undergoes an unusual zero energy motion - in its
effect reminiscent of a limited rotational hinge.
\end{abstract}


\affiliation{$^{1}$Groupe BioPhysStat, Universit\'{e} Paul
Verlaine, 57078 Metz, France
\\ $^{2}$ CNRS, Institut Charles Sadron, 23 rue du Loess BP 84047, 67034
Strasbourg, France }

 \maketitle

Microtubules are the stiffest cytoskeletal component and play versatile and
indispensable roles in living cells. They act as cellular bones, transport
roads \cite{Genref MTs} and cytoplasmic stirring rods \cite{MTStirringRod}.
Microtubules consist of elementary building blocks - the tubulin dimers - that
polymerize head to tail into linear protofilaments (PFs). PFs\ themselves
associate side by side to form the hollow tube structure known as the
microtubule (MT). Despite a long history of their biophysical study a deeper
understanding of MT's elastic and dynamic properties remains elusive to this
date. Besides the unusual polymerization related non-equilibrium features like
''treadmilling''\ and the dynamic instability there are a number of other
experimental mysteries - in thermal equilibrium- that presently defy coherent
explanations, most notably: (i) The presence of high ''intrinsic
curvature''\ \cite{Venier}-\cite{Janson} of unclear origin, also identified as
a long wave-length helicity \cite{Venier}. (ii)\ In various active bending
\cite{ActtiveMTBending} or thermal fluctuation experiments \cite{Pampaloni}%
\cite{Taute}\cite{Keller} MTs display length dependent, even non-monotonic
apparent stiffness \cite{Taute}. (iii) They exhibit unusually slow thermal
dynamics in comparison with standard semiflexible filaments \cite{Taute}%
\cite{Janson}.

\begin{figure}[ptb]
\includegraphics*[width=8cm]{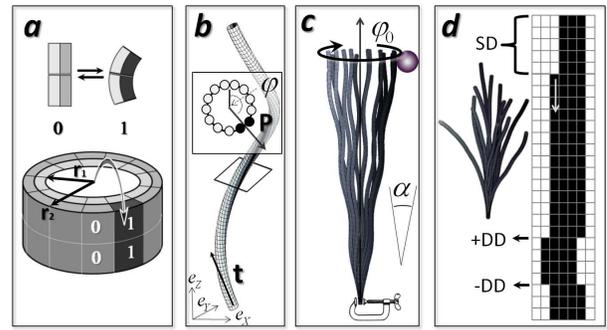}\caption{Polymorphic Tube Model:
(a)\ The tubulin dimer fluctuates between two states $\sigma=0,1$ (straight /
curved), b)\ Tubulin switching on one MT side leads to spontaneous breaking of
symmetry. Combined with the built-in lattice twist MT\ forms a polymorphic
helix. The polymorphic order parameter $P$\ with phase angle $\phi$ describes
the distribution of tubulin states in the cross section. c)\ Polymorphic
wobbling - the zero-energy motion of the phase angle at each cross-section. It
is responsible for the rotation on a cone with opening angle $\alpha$ when
clamped by the end. d) Defects in polymorphic order: Single defects\ (SD) have
a cost proportional to their length. Double defects (DD) give only local
energy contribution.}%
\label{fig1}%
\end{figure}

The most bizarre and controversial feature (ii) has been the
subject of much debate and some theoretical explanation attempts
based on low shear stiffness modulus have been put forward
\cite{Frey}. However a careful reanalysis of clamped MT
experiments, Figs $2,3$ reveals two features not captured by these
initial models: the lateral end-fluctuations scale as $\sim L^{2}$
while the relaxation times scale as $\sim L^{3}$. This exotic
behavior naively suggests the presence of a limited angular hinge
at the MT\ clamping point. On the other hand artifacts that could
trivially lead to a ''hinged behavior''\ (like loose MT\
attachment and punctual MT\ damage) were specifically excluded in
experiments \cite{Pampaloni}\cite{Taute}. We will outline here a
model based on \textit{internal} MT dynamics explaining phenomena
(i)-(iii). It leads us to the origin of MT helicity (i) implying
(ii)-(iii) as most natural corollaries \cite{CommentTax}. The two
central assumptions of our model are as follows: (I) The tubulin
dimer is a conformationally multistable entity and fluctuates
between at least 2 states on experimental time scales. (II) There
is a nearest-neighbor cooperative interaction of tubulin states
along the PF axis. We are lead to assumptions I-II from several
independent directions: \textit{First}, the experimentally
observed MT helicity \cite{Venier} implies that there is a
symmetry breaking mechanism of individual PF's conformational
properties. In analogy to the classic case of bacterial flagellum
the existence of helices in azimuthaly symmetric bundles also
necessitates a cooperative longitudinal interaction along
protofilaments \cite{Calladine}\cite{Powers}. \textit{Second},
investigations of single protofilament conformations by
Elie-Caille et al \cite{Multistable Tub EM} reveal that a single
taxol- PF can coexist in at least 2 states with comparable free
energy: a straight state $\kappa _{PF}\approx0$ and a weakly
curved state with intrinsic curvature $\kappa
_{PF}\approx1/250nm$. These authors also point out the apparent
cooperative nature of straight to curved transition within single
PFs. \textit{Third}, when mechanically buckled by AFM\ tips
tubulin dimers occasionally switch back to the initial straight
conformation \cite{MTIntendation}. \textit{Fourth}, tubulin
multistability was inferred from the formation of stable circular
MT arcs in kinesin driven gliding assays by Amos \& Amos
\cite{Amos}. Unfortunately their clear, seminal observations were
subsequently forgotten for decades leading to much of the
confusion about MTs we are witnessing today.

\textit{Polymorphic MT Model}. Starting from assumptions $I-II$ we model the
tubulin dimer state by a two state variable $\sigma_{n}\left(  s\right)  =0,1$
(the tubulin dimer in the ''straight''/''curved''\ state, cf. Fig. 1a) at each
lattice site with circumferential PF index $n=1,...N$ ($N=11-16$ number of
PFs) at longitudinal arclength centerline position $s$. The total elastic
+\ conformational energy can be written as $E_{MT}=\int\nolimits_{0}%
^{L}\left(  e_{el}+e_{trans}+e_{inter}\right)  ds$ with%

\begin{align}
e_{el}  &  =\frac{Y}{2}\int\int\left(  \varepsilon-\varepsilon_{pol}\right)
^{2}rdrd\alpha\label{Ener1}\\
e_{trans}  &  =-\frac{\Delta G}{b}\sum\nolimits_{n=1}^{N}\sigma_{n}%
(s)\text{,}\label{Ener2}\\
e_{inter}  &  =-\frac{J}{b}\sum\nolimits_{n=1}^{N}\left(  2\sigma_{n}\left(
s\right)  -1\right)  \left(  2\sigma_{n}\left(  s+b\right)  -1\right)
\label{Ener3}%
\end{align}
where the integration in $e_{el}$ goes over the annular MT cross-section with
$r_{1}\approx7.5nm,$ $r_{2}\approx11.5nm$ the inner and outer MT\ radii, with
$\Delta G>0$ the energy difference between the $0$ and $1$ state
and$\ b\approx8nm$ the monomer length, $J$ the ''Ising''\ cooperative coupling
term along the PF contour and with the polymorphism induced prestrain
$\varepsilon_{pol}\propto\varepsilon_{PF}\sigma_{n}\left(  s\right)  $
\cite{EpsilonPol} where $\varepsilon_{PF}$ is the strain generated in the
curved state. The latter can be estimated from the switched PF curvature
$\kappa_{PF}\approx\left(  250nm\right)  ^{-1}$ \cite{Multistable Tub EM} to
be $\varepsilon_{PF}=d_{PF}\kappa_{PF}/2\approx10^{-2}$. For an isotropic
Euler-Kirchhoff beam, the actual material deformations are related to the
centerline curvature via $\varepsilon=-\vec{\kappa}\cdot\vec{r}$ with $\vec
{r}$ the radial vector in the cross-section$.$

Upon inspection it becomes clear that the phase behavior (straight or curved
state stability)\ is contained in the interplay of the first two terms
$e_{el}$ and $e_{trans}$ while the thermal dynamics is governed by the $3$rd
$e_{inter}$ which rules over defect behavior (cf. Fig. 1d). To understand the
basic behavior we first consider a short MT section$\;$along which the PFs are
in a uniform state $\sigma_{n}(s)=\sigma_{n}(s+b)$ ($e_{inter}=$const. can be
dropped). Furthermore we resort to the single block ansatz, i.e. at each
cross-section there is only one continuous block of switched PFs of length
$p$. This ansatz was successfully used by Calladine in modelling bacterial
flagellin polymorphic states \cite{Calladine}. In this approximation the
energy density becomes
\begin{equation}
e=\frac{B}{2}\left(  \left(  \kappa-\kappa_{pol}(p)\right)  ^{2}+\kappa
_{1}^{2}\left(  \gamma\frac{\pi}{N}p-\sin^{2}\left(  \frac{\pi}{N}p\right)
\right)  \right)  \label{EnergyL}%
\end{equation}
with the bending modulus $B=\frac{Y\pi}{4}\left(  r_{2}^{4}-r_{1}^{4}\right)
$ and the polymorphic curvature $\kappa_{pol}(p)=\kappa_{1}\sin\left(
\frac{\pi}{N}p\right)  $ with $\kappa_{1}=\frac{\kappa_{FP}\left(  r_{2}%
-r_{1}\right)  ^{2}}{\pi\left(  r_{1}^{2}+r_{2}^{2}\right)  }$. The MT phase
behavior depends on the polymorphic-elastic competition parameter
$\gamma=\frac{\kappa_{PF}}{\kappa_{1}}-\frac{2N\Delta G}{bB\kappa_{1}^{2}}.$
Physically, $\gamma$ measures the ratio between polymorphic energy of tubulin
switching and the elastic cost of the transition. For $\gamma<-1$ all the PFs
are in the (highly prestrained) state $\sigma=1$ while for $\gamma>1$ all them
are in the state $\sigma=0$ - both corresponding to a straight MT. For
$-1<\gamma<1$ we have coexistence of $2$ locally (meta) stable states
:\ straight ($p=0$ or $p=N$)\ and curved state with $p>0$. For $-\overline
{\gamma}<\gamma<\overline{\gamma}$ with $\overline{\gamma}\approx0.72$ the
curved state is the absolute energy minimum and the straight state is only
metastable. Therefore in this regime, the ground state of a microtubule
bearing natural lattice twist will be helical (cf. Fig. 1b). Assuming a stable
helical state as observed in \cite{Venier} $\ $we have $p/N\in\lbrack1/4,1/2]$
giving us an estimate for the radius of curvature $\kappa_{pol}^{-1}%
\approx9-14\mu m$. This compares favorably with an estimate of observed
helices $\kappa^{-1}\approx11\mu m$ from \cite{Venier}. The helical stability
and the magnitude of the protofilament curvature $\kappa_{PF}\approx1/250nm$
\cite{Multistable Tub EM} with a typical protein Young modulus $Y\approx
1-10GPa,$ allows us also a simple estimate of the transition energy per
monomer $\Delta G\approx+1.1$ to $+11kT.$ In general, the energy in
Eqs.\ref{Ener1}-\ref{Ener3}\ gives rise to a complex behavior and we focus on
basic phenomena. It turns out that a most remarkable deviation from standard
wormlike chain behavior comes from the change of polymorphic phase that we
consider in the following.

\textit{Polymorphic Phase Dynamics. }To better understand the
central phenomenon, we define at each MT cross-section the complex
\textit{polymorphic order parameter} $P\left(  s\right)
=\sum_{n=1}^{N}e^{2\pi in/N}\sigma _{n}\left(  s\right)  =\left|
P\left(  s\right)  \right|  e^{i\phi\left( s\right)  }$ where
$\left|  P\left(  s\right)  \right|  $ denotes the ''polymorphic
modulus''\ and $\phi$ the ''polymorphic phase''\ (cf. Fig. 1b).
The polymorphic state can then be described by the local (complex)
centerline curvature
$\hat{\kappa}_{pol}(s)=\kappa_{0}e^{iq_{0}s}P(s)$ with $\kappa
_{0}=\kappa_{1}\sin\pi/N$ and $q_{0}$ the natural lattice twist
that varies with PF number \cite{Kinesin Rotation}. This gives
rise to a helical MT shape described by the curvature $\left|
\hat{\kappa}_{pol}\right|  =\kappa_{pol}$ and torsion
$\tau\approx\phi^{\prime}+q_{0}$. For large acting forces both the
polymorphic phase $\phi$ and amplitude $\left|  P\right|  $ will
vary along the contour, however for small (thermal) perturbations
the phase fluctuations will be dominant \cite{Phase Footnote}.
Based on this and on the observation of stable helical states
\cite{Venier} we will now assume $\left|  P\right| =const,$ and
write the total energy of the MT\ whose centerline deflection is
described by a complex angle $\theta\left(  s\right)
=\theta_{x}\left( s\right)  +i\theta_{y}\left(  s\right)  $
(deflection angles in x/y direction) as follows:\
\begin{equation}
E_{tot}=E_{el}\left(  \theta,\phi\right)  +E_{pol}\left(  \phi\right)
\label{Etotal}%
\end{equation}
The first energy term is the ''wormlike-chain''\ bending contribution
$E_{el}\left(  \theta,\phi\right)  =\frac{B}{2}\int\left|  \theta^{\prime
}-\hat{\kappa}_{pol}\right|  ^{2}ds$. The second term is the polymorphic phase
energy $E_{pol}\left(  \phi\right)  =\frac{C_{\phi}}{2}\int_{0}^{L}%
\phi^{\prime2}ds$ with the polymorphic phase stiffness $C_{\phi}=k_{B}%
T\frac{N^{2}b}{8\pi^{2}}\left(  2+e^{2J/k_{B}T}\right)  $ which can be related
to the density of double defects with energy $2J\;$(cf. Fig. 1d), giving rise
to a new length scale - the polymorphic phase coherence length $l_{\phi
}=C_{\phi}/k_{B}T.$

\begin{figure}[ptb]
\includegraphics*[width=6.5cm]{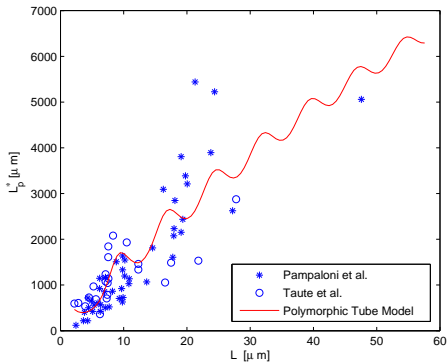}\caption{Effective persistence length
$l_{p}^{\ast}\left(  s\right)  $ as a function of the position
from the attachment point along the MT contour. The experimental
and theoretical prediction with $l_{B}=25mm,$ $\lambda=7.5\mu m,$
$\kappa_{0}^{-1}=18\mu m,$
$q_{0}l_{\phi}>>1$.}%
\label{fig2}%
\end{figure}\ The most unusual property of a polymorphic chain is reflected in
the rotational invariance of $E_{pol}\left(  \phi\right)  $. The
broken cylindrical to helical symmetry of the straight state is
restored by the presence of a ''Goldstone mode''
$\phi\rightarrow\phi+\phi_{0}$ \cite{CommentContiuity} consisting
of a rotation of $P$ by an arbitrary angle $\phi_{0}$ in the
material frame (cf. Fig. 1c). This mode that we will call the
''wobbling mode''\ is a fundamental property of a helically
polymorphic filament. The wobbling mode leads to dramatic effects
on chain's fluctuations and is the clue to the resolution of
mysteries (i)-(iii). To see this we will first investigate the
static properties resulting in length dependent variations of the
persistence length.

\textit{Persistence Length Anomalies.} Among several definitions of the
persistence length \cite{Persistence Footnote} we consider for direct
comparison with experiments \cite{Pampaloni}\cite{Taute}, the lateral
fluctuation persistence length $l_{p}^{\ast}(s)=\left(  2/3\right)
s^{3}/\left\langle \left|  \rho\left(  s\right)  \right|  ^{2}\right\rangle $
with $\rho\left(  s\right)  =x(s)+iy(s)$ the transverse displacement at
position $s$ of a MT clamped at $s=0$ and $\left\langle ..\right\rangle $ the
statistical average. It is easy to see from Eq. \ref{Etotal} that for small
deflections, $\rho$ decouples into independent elastic and polymorphic
displacements $\rho\left(  s\right)  =\rho_{el}+\rho_{pol}$, such that
$l_{p}^{\ast}=\left(  l_{pol}^{\ast-1}+l_{B}^{-1}\right)  ^{-1}$ with
$l_{pol}^{\ast}=\left(  2/3\right)  s^{3}/\left\langle \left|  \rho
_{pol}\right|  ^{2}\right\rangle $ where $\rho_{pol}=\kappa_{0}\int_{0}%
^{s}\int_{0}^{s_{1}}e^{iq_{0}\tilde{s}+i\phi\left(  \tilde{s}\right)
}d\widetilde{s}ds_{1}$. The coherent helix nature of the MT observed in
\cite{Venier} and the absence of a plateau in $l_{p}^{\ast}$ imply
\cite{Plateau Footnote} that $l_{\phi}>>\lambda=2\pi q_{0}^{-1}$ (the helix
wave length). In that limit we obtain $\left\langle \left|  \rho_{pol}\right|
^{2}\right\rangle \approx\kappa_{0}^{2}q_{0}^{-2}[\frac{2}{q_{0}^{2}}%
+s^{2}+\frac{s^{3}}{3l_{\phi}}-\frac{2}{q_{0}^{2}}e^{-\frac{s}{2l_{\phi}}%
}((1+\frac{s}{2l_{\phi}})\cos q_{0}s+sq_{0}\sin q_{0}s)]$. \begin{figure}[ptb]
\includegraphics*[width=6cm]{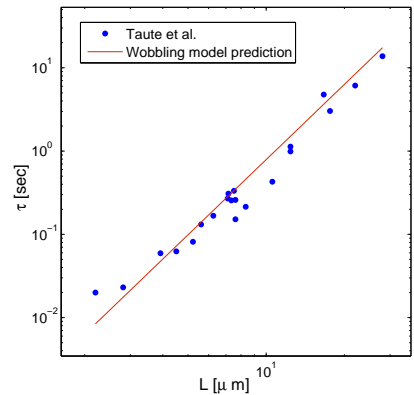}\caption{Longest relaxation time of a
microtubule of length L, experimental data from \cite{Taute} and theoretical
prediction from the wobbling mode approximation and the wobbling angle
$\alpha=2\kappa_{0}/q_{0}\approx7.\,\allowbreak9^{\circ}$ extracted from
static data \cite{Pampaloni}\cite{Taute}, Fig. 2. }%
\label{fig3}%
\end{figure}Whereas for an ideal WLC $l_{p}^{\ast}=l_{B}\equiv B/kT$ is
position and definition independent, the polymorphic fluctuations
induce a strong position / distance dependence - a behavior that
could be interpreted as ''length dependent persistence length''.
Indeed for $l_{\phi}>>$ $s>>q_{0}^{-1}$ the persistence length
displays a non-monotonic oscillatory behavior around a nearly
linearly growing average value $l_{p}^{\ast}\left(
s\right)  \approx\frac{2}{3}\frac{q_{0}^{2}}{\kappa_{0}^{2}}s+\frac{4}{3}%
\frac{q_{0}}{\kappa_{0}^{2}}\sin\left(  q_{0}s\right)  $. This oscillation is
related to the helical ground state while the linear growth $l_{p}^{\ast
}\left(  s\right)  \propto\alpha^{2}s$ is associated to the conical rotation
of the clamped chain (wobbling mode), cf. Fig. 1c, with an angle
$\alpha=2\kappa_{0}q_{0}^{-1}$. For $s>>l_{\phi}$ the saturation regime with a
renormalized $l_{p}^{\ast}\left(  \infty\right)  =1/\left(  l_{pol}^{-1}%
+l_{B}^{-1}\right)  $\ with
$l_{pol}=\allowbreak2l_{\phi}q_{0}^{2}\kappa _{0}^{-2}$ is
reached. The theory can now be compared with the experimental data
\cite{Pampaloni}\cite{Taute} (cf. Fig. 2) that reveal several
interesting characteristics in agreement with predictions. In
particular the mean linear growth of $l_{p}^{\ast}\left(  L\right)
$ (single parameter fit $l_{p}^{\ast }\sim L^{\delta}$ gives
$\delta=1.05$) and the non-monotonic $l_{p}^{\ast}\left( L\right)
$ dependence \cite{Taute} are well captured by the theory. The
linearly growing experimental spread of $l_{p}^{\ast}$ with $L$ is
likely linked to the spread of $q_{0}$ in the MT lattice
populations \cite{Kinesin Rotation}. The large length plateau
$s>>l_{\phi}$ is not reached even for longest MTs ($\sim50\mu m$)
in agreement with coherent helices \cite{Venier}. Our best
comparison between theory and experiments (cf. Fig. 2) gives
$l_{B}=25mm$ corresponding to $Y\approx9GPa$ (proteins with $Y\
$up to $19GPa$ exist \cite{15 GPa Tubes}) and a helix wave length
$\lambda\approx7.5\mu m.$ This is close to the expected $6\mu m$
corresponding to the twist \cite{Kinesin Rotation} of the
predominant $14$ PF MTs fraction in the in-vitro MTs preparation
of \cite{Pampaloni}\cite{Taute}. It turns out that $l_{B}$ is
larger than in previous studies $l_{B}\sim1-6mm$ where however
polymorphic fluctuations were neglected. The absence of the
plateau also allows a lower estimate of the coherence length
$l_{\phi}>55\mu m$ and the coupling constant $J>4k_{B}T$.

\textit{Polymorphic Phase Dynamics. }To describe the MT fluctuation dynamics
we consider the total dissipation functional $P_{diss}=P_{ext}+P_{int}$ which
is composed of an internal dissipation contribution $P_{int}=\frac{1}{2}%
\xi_{int}\int\dot{\phi}^{2}ds$ and an external hydrodynamic\ dissipation
$P_{ext}=\frac{1}{2}\xi_{\perp}\int\left|  \dot{\rho}\right|  ^{2}ds$ with
$\xi_{\perp}=4\pi\eta/\left(  \ln\left(  2L/r\right)  -1/2\right)  $ the
lateral friction constant, $\eta$ the solvent viscosity, $r$ and $L\;$the MT
radius and length. The time evolution equation of the phase variable
$\phi\left(  s,t\right)  $ and elastic displacement $\rho_{el}\left(
s,t\right)  $ is given by the coupled Langevin equations $\frac{\delta
E}{\delta\phi}=-\frac{\delta P_{diss}}{\delta\dot{\phi}}+\Gamma_{\phi}$ and
$\frac{\delta E}{\delta\rho_{el}}=-\frac{\delta P_{diss}}{\delta\dot{\rho
}_{el}}+\Gamma_{\rho}$ with $\Gamma_{\phi/\rho}$ the thermal noise term. In
general this dynamics is highly non-linear however in the experimentally
relevant regime where the behavior is dominated by the wobbling mode the
equations simplify greatly and we end up with a simple diffusive behavior of
the wobbling mode $\frac{d}{dt}\phi_{0}\left(  t\right)  =\frac{1}{\xi_{tot}%
}L^{-1}\int_{0}^{L}\Gamma_{\phi}\left(  s,t\right)  ds$ with a friction
constant given by $\xi_{tot}=\xi_{int}+\xi_{ext}$ where $\xi_{ext}=2\xi
_{\perp}\kappa_{0}^{2}q_{0}^{-4}(\left(  1+\cos Lq_{0}\right)  -4\sin
Lq_{0}+q_{0}^{3}L^{3}/3).$ For comparison with the experiment we compute the
time correlation of the $y$ deflection. A short calculation gives
$\left\langle y_{pol}(L,t)y_{pol}(L,t^{\prime})\right\rangle \varpropto
e^{-\left|  t-t^{\prime}\right|  /\tau\left(  L\right)  }$ with the relaxation
time $\tau\left(  L\right)  \approx L\xi_{tot}/k_{B}T$. For small lengths,
$\tau\left(  L\right)  \approx L\xi_{int}/k_{B}T$ is dominated by internal
dissipation while for large lengths $\tau\left(  L\right)  \approx\frac
{\xi_{\perp}}{3k_{B}T}\left(  \kappa_{0}/q_{0}\right)  ^{2}L^{3}.$ A careful
analysis of the experimental data \cite{Taute} reveals in fact the latter
scaling. An independent single exponent fit gives $\tau\propto L^{\alpha}$
with $\alpha=2.9$. Using the value $\left(  \kappa_{0}/q_{0}\right)
^{2}\approx4.\,8\times10^{-3}$ from Fig. 2 and $\xi_{\perp}\approx2\eta$ with
$\eta=10^{-3}Pa\cdot s$ \cite{Time Footnote} we find the theoretical value
$\tau_{th}/L^{3}=7.9\times10^{14}s/m^{3}$ that can be compared with the fit of
experimental data (Fig. 3)\ $\tau_{fit}/L^{3}=6.\,25\times10^{14}s/m^{3}$. The
excellent agreement of both the exponent and the prefactor leads us again to
the strong conclusion that in these experiments the clamped MT is an almost
rigid helical polymorphic rotor whose behavior is dominated by the zero energy
(''wobbling'')\ mode and hydrodynamic dissipation. For very short MTs the
linearly scaling internal dissipation dominates and we could measure
$\xi_{int}$ from the limit value of $\tau_{th}/L,$  for $L\rightarrow0$. For
the available data $L>2$ $\mu m$ \cite{Taute} this plateau-regime is not yet
fully developed and we can only provide an upper estimate from the data
$\xi_{int}\lesssim4\times10^{-17}Ns$.

\textit{Conclusion}. The MT fluctuations are well described - both dynamically
and statically - by the bistable tubulin model and the reason for appearance
of MT helices becomes obvious. The otherwise mysterious lateral fluctuations
reflected in $l_{p}^{\ast}\left(  L\right)  \sim L^{1}$\ and $\tau_{p}\left(
L\right)  \sim L^{3}$ scaling\ are mere consequences of the ''wobbling
motion''\ of a polymorphic cooperatively switching helical lattice. We
speculate that the implied conformational multistablity of tubulin and the
allosteric interaction are not just nature's way to modulate the elastic
properties of its most important cytoskeletal mechano-element. It could also
be a missing piece in the puzzle of dynamic instability. Another intriguing
possibility of using this switch for long range conformational signalling in
vivo, could hardly have been overlooked by evolution. I.M.K thanks Francesco
Pampaloni for stimulating discussions.


\begin{thebibliography}{99}
\bibitem {CommentTax}We will focus on the simplest case of equilibrated, taxol
stabilized MTs where dynamic instability is absent yet the very rich thermal
behavior i-iii is pronounced.
\bibitem {EpsilonPol}$\varepsilon_{pol}=\varepsilon_{PF}\sigma_{n}\left(
s\right)  \left[  I_{[R-2d_{PF},R-d_{PF}]}\left(  r\right)  -I_{[R-d_{PF}%
,R]}\left(  r\right)  \right]  $ $\cdot I_{[\frac{2\pi}{N}n+q_{0}s,\frac{2\pi
}{N}\left(  n+1\right)  +q_{0}s]}\left(  \alpha\right)  $ where $I_{[.]}%
\left(  x\right)  =1$ if $x\in\lbrack.]$ and $0$ otherwise, $d_{PF}$ the PF
diameter and $q_{0}$ the natural lattice twist.
\bibitem {Phase Footnote}Phase $\phi$ fluctuations are induced by double
defects which carry a limited local energy cost $\Delta E=2Jn$ ($n$ being the
number of double defects). $\left|  P\right|  $ variations are induced by
single defects having in general larger energy cost that grows with their end
distance, cf. Fig.1d.
\bibitem {CommentContiuity}This mode has a N fold symmetry however for large
number of PFs\ $N=11-15$ it can be approximated as contiuuous. \
\bibitem {Persistence Footnote}Another more common definition, from angular
correlation $\left\langle \cos\left(  \theta\left(  s\right)  -\theta\left(
s^{\prime}\right)  \right)  \right\rangle \ $exhibiting a similarly rich
behavior as $l_{p}^{\ast}$ (yet a distinct functional form) will be discussed elsewhere.
\bibitem {Plateau Footnote}In the plateau region the helix looses its
''coherence''\ and the collective rigid rotational conical motion softens
until an uncorrelated segment movement becomes dominant.
\bibitem {Time Footnote}In the experimental range \cite{Taute} of $2.2\mu
m<L<28\mu m$ : $\xi_{\perp}\approx1.6\eta-2.3\eta$ is almost length independent.
\end{thebibliography}
\end{document}